\documentclass[12pt]{iopart}
\usepackage{graphicx}
\eqnobysec

\begin{document}

%%%%%%%%%%%%%%%%%%%%%%%%%%%%%%%%%%%%%%%%%%%%%%%%%%%%%%%%%%%

\title[Expressions of the Green function]
{Expressions of the Green function in terms of reflection coefficients}

\author{Toru Miyazawa}

\address{Department of Physics, Gakushuin University, 
Tokyo 171-8588, Japan}
\ead{toru.miyazawa@gakushuin.ac.jp}
\begin{abstract}
We study the one-dimensional Schr\"odinger equation and derive
exact expressions for the Green function in terms of reflection coefficients which are defined for semi-infinite intervals.
We also discuss the relation between our results and the WKB approximation.
\end{abstract}

\maketitle

%%%%%%%%%%%%%%%%%%%%%%%%%%%%%%%%%%%%%%%%%%%%%%%%%%%%%%%%%%%

\section{Introduction}

Let us consider the Green function for the steady-state Schr\"odinger equation in one dimension,
\begin{equation}
\label{1-1.1}
-\frac{d^2}{dx^2}\psi(x)+V_{\rm S}(x)\psi(x)=k^2\psi(x).
\end{equation}
The Green function plays important roles in various physical problems, and there are many approaches to the study of the Green function. 
In this paper we discuss a new description of the Green function in terms of reflection coefficients.

From the physical point of view, it is natural to interpret the propagation of waves in terms of the processes of multiple reflections and transmissions.
In quantum mechanics, this interpretation has been used, for the most part, in the context of semiclassical approximations \cite{landauer, kira}.
The Bremmer series, which is a perturbative improvement of the WKB approximation, is based on this picture \cite{bremmer}. 
A similar idea is used in the invariant imbedding method \cite{bellman}, which has applications in many areas including transport problems in astrophysics, conductors, and random media [5--8].
The essence of the invariant imbedding method is to express everything in terms of ^^ ^^ emergent" or ^^ ^^ observable" quantities such as transmission and reflection coefficients, without need of considering what is happening within the system. 
In this method one deals with reflection coefficients for finite intervals, and, by varying the endpoint of the interval, derives a differential equation of Riccati type satisfied by the reflection coefficients. 
The derivation of this Riccati equation is essentially equivalent to taking account of the transmission and reflection processes at the endpoint. 

It is possible to use the same idea to construct the Green function. By taking the sum over all the multiple reflections and transmissions, we can derive  exact expressions for the Green function~\cite{theory, luz}\footnote{
The expression for the Green function derived in \cite{luz} for segmented potentials is identical to the one obtained in \cite{theory} for the Fokker-Planck equation.
}.
These expressions are written in terms of the transmission coefficient for a finite interval, and the reflection coefficients for finite and semi-infinite intervals. 
 With these expressions, the analysis of the Green function can be reduced to that of the transmission and reflection coefficients. 

The structure of reflection coefficients for semi-infinite intervals have been throughly studied, and various formulas have been obtained for their high- and low-energy  behaviors \cite{analysis}.
However, the mathematical structure of transmission coefficients is not as simple. This is because transmission coefficients are ^^ ^^ non-local" quantities in the sense that they are functions of two endpoints of the finite interval. 
We may say that, in a sense, reflection coefficients are more fundamental quantities than transmission coefficients.
The analysis of the Green function becomes much easier if it is expressed solely in terms of reflection coefficients for semi-infinte intervals, without using transmission coefficients. 
It is the objective of the present paper to derive such expressions.   

The expressions in terms of reflection coefficients are particularly useful for the analysis in the high- and low-energy regions. 
By using the formulas already known for the reflection coefficients, we can derive new formulas for asymptotic expansions of the Green function. The advantage of this approach over conventional methods is that it can be applied to a larger class of potentials. The reflection coefficients can be defined irrespective of whether the potential $V_{\rm S}(x)$ is finite or infinite as $x \to \pm \infty$; we do not need to assume that $V_{\rm S}(x)$ vanishes sufficiently rapidly at infinity, as is necessary for the description using Jost solutions. We do not need to care about the existence of bound states, nor do we have to know the eigenvalues. Conventional methods of analysis are sensitive to the behavior of the potential at infinity, and it is often necessary to use different methods for different kinds of potentials. In the formulation in terms of reflection coefficients, the analysis of the Green function can be carried out for various types of potentials in a unified way. 
In addition, the formulas for asymptotic expansions obtained in this method are more explicit than the ones obtained by conventional methods. (This will be discussed in a separate paper.)  

The expressions in terms of reflection coefficients are also convenient for calculating the Green function in practical situations, either approximately or numerically.
It turns out that the expressions derived in this paper have a close relation with the WKB method. 
In the light of the formalism developed here, we can understand the WKB method from a new viewpoint, which may possibly lead to new improvements of the WKB approximation.
Our expressions can also be used as a basis for other new approximation methods.
Since the reflection coefficients are quantities that have a clear physical meaning, expressing the Green function in terms of them is useful for the purpose of making approximations.
The reflection coefficients are also suited for numerical treatments, and so these expressions will be useful for the numerical calculation of the Green function, too.

In our method of derivation, we make use of the Fokker-Planck equation.
It is well known that the Schr\"odinger equation (\ref{1-1.1}), with an appropriate shift of the energy level, can be transformed into a Fokker-Planck equation~\cite{risken}.
The (time-independent) Fokker-Planck equation describing the Brownian motion in a potential $V(x)$ has the form
\begin{equation}
\label{1-1.2}
-\frac{d^2}{dx^2}\phi(x)+2 \frac{d}{dx}[f(x)\phi(x)]=k^2\phi(x),
\end{equation}
where
\begin{equation}
\label{1-1.3}
f(x)\equiv -\frac{1}{2}\frac{d}{dx}V(x).
\end{equation}
Equation (\ref{1-1.2}) is equivalent to (\ref{1-1.1}), where
\begin{equation}
\label{1-1.4}
\psi(x)=e^{V(x)/2}\phi(x), \qquad V_{\rm S}(x)=f'(x)+f^2(x).
\end{equation} 
With the use of the Fokker-Planck equation, it becomes easier to study the structure of the transmission and reflection coefficients, and various formulas take simpler forms. 
In particular, a symmetry transformation of the Fokker-Planck equation plays a crucial role in our method. 
As a result we obtain a one-parameter family of expressions, which reflects the symmetry structure of the Fokker-Planck equation.

We assume that $V_{\rm S}(x)$ either converges to a finite value or diverges to $+\infty$ as $x \to +\infty$, and that $V_{\rm S}(x)$ is also either finite or $+\infty$ as $x \to -\infty$.
(We do not consider the cases where $V_{\rm S}(x)$ tends to $-\infty$ as $x \to \pm \infty$, or the cases where $V_{\rm S}(x)$ oscillates at infinity.) 
We also assume that $k$ is, in general, a complex number with ${\rm Im}\,k \geq 0$.
Let $G_{\rm S}(x,x';k)$ denote the Green function for equation~(\ref{1-1.1}), satisfying
\begin{equation}
\label{1-2.1}
\left[
\frac{\partial^2}{\partial x^2}-V_{\rm S}(x)+k^2
\right]
G_{\rm S}(x,x';k)=\delta(x-x')
\end{equation}
with the boundary condition $G_{\rm S}(x,x';k) \to 0$ as $\vert x-x'\vert \to \infty$ for ${\rm Im}\,k>0$.
We define
\begin{equation}
\label{1-2.2}
G(x,x';k)\equiv 2ik G_{\rm S}(x,x';k).
\end{equation}
In this paper we shall deal with the quantity $G$ defined by (\ref{1-2.2}), rather than $G_{\rm S}$ itself.
(For convenience, we shall also call $G$ the Green function.) 
Without loss of generality we may assume that $x \geq x'$. The expressions for $x<x'$ are obtained by interchanging $x$ and $x'$.

Let us now define the reflection coefficients for semi-infinite intervals. 
For general $V_{\rm S}(x)$ (rather than special forms such as piecewise constant or segmented potentials), there is no unique natural way of defining the reflection coefficients for finite or semi-infinite intervals. 
As mentioned above, we shall define them in terms of the Fokker-Planck equation, and this turns out to give the simplest description.
Our definition of the reflection coefficients for semi-infinte intervals is illustrated in figure~1.
\begin{figure}
\hspace{2cm}
\includegraphics[scale=0.6]{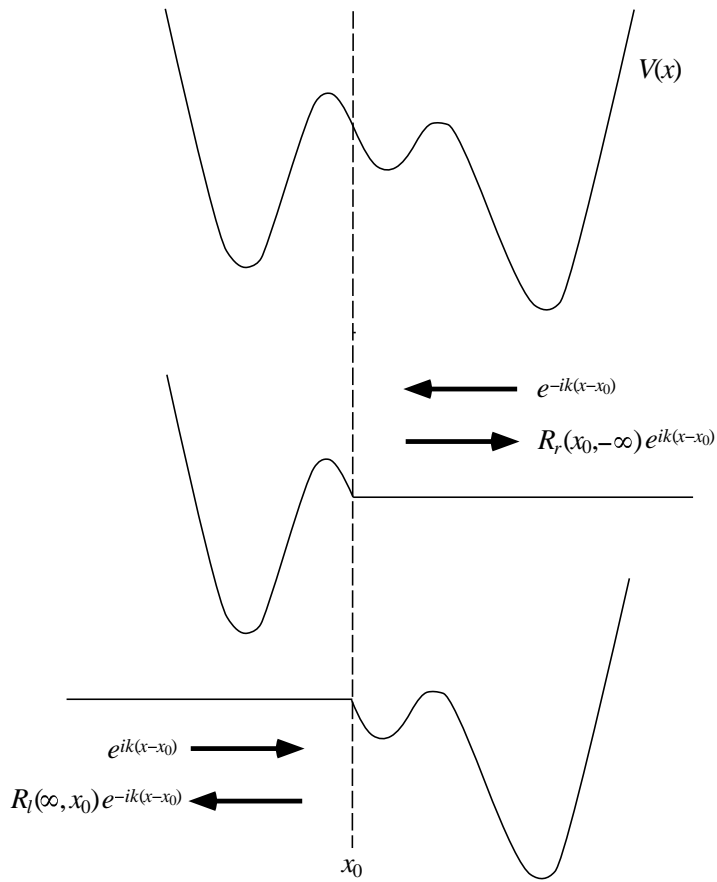}
\caption{
Definition of $R_r(x_0,-\infty;k)$ and $R_l(\infty,x_0;k)$.
}
\end{figure}
Let $x_0$ be an arbitrarily chosen point. 
 We let the Fokker-Planck potential $V(x)$ in the region $x>x_0$ be replaced by the constant value $V(x_0)$, and define
\begin{equation}
\label{1-1.5}
\bar V(x)\equiv 
V(x) \theta(x_0-x) +V(x_0) \theta(x-x_0),
\end{equation}
where $\theta$ is the Heaviside step function. 
(Recall that the Schr\"odinger potential $V_{\rm S}(x)$ is related to $V(x)$ by equations (\ref{1-1.3}) and (\ref{1-1.4}).) 
We consider equation~(\ref{1-1.2}) with $f(x)$ replaced by $\bar  f(x) \equiv -(1/2)(d/dx)\bar V(x)$. In the region $x>x_0$, where $\bar f(x)=0$, this equation has independent solutions of the form $e^{+ikx}$ and $e^{-ikx}$.  We define the reflection coefficient $R_r(x_0,-\infty;k)$ as the coefficient multiplying the reflected wave $e^{ik(x-x_0)}$ in the region $x>x_0$ when there is an incident wave $e^{-ik(x-x_0)}$. 
In other words, $R_r(x_0,-\infty;k)$ is defined by a solution of the form
\refstepcounter{equation}
\label{1-1.6}
\addtocounter{equation}{-1}
\numparts
\begin{eqnarray}
\label{1-1.6a}
\phi(x)&=e^{-ik(x-x_0)} + R_r(x_0,-\infty;k) e^{ik(x-x_0)} \quad {\rm for} \quad x>x_0,
\\
\label{1-1.6b}
\phi(x)& \to 0 \quad {\rm as} \quad x \to -\infty. 
\end{eqnarray}
\endnumparts
(When $k$ is real, it is necessary to assume in (\ref{1-1.6b}) that $k$ has an infinitesimal imaginary part $i \epsilon$ with $\epsilon>0$.) 
If $V_{\rm S}(x)=0$ for $x_1<x$ with some $x_1$, and if $x_1<x_0$, then the above definition of $R_r(x_0,-\infty;k)$ coincides with the usual definition of the reflection coefficient. In general cases, the Schr\"odinger potential corresponding to the Fokker-Planck potential (\ref{1-1.5}) includes a delta function at $x=x_0$. 
In the same way, the left reflection coefficient for the interval $(x_0,+\infty)$ is defined by considering, instead of (\ref{1-1.5}) and (\ref{1-1.6}),
\begin{equation}
\label{1-7}
\bar V(x)\equiv 
V(x) \theta(x-x_0) +V(x_0) \theta(x_0-x),
\end{equation}
and
\refstepcounter{equation}
\label{1-1.8}
\addtocounter{equation}{-1}
\numparts
\begin{eqnarray}
\label{1-1.8a}
\phi(x)&=e^{ik(x-x_0)} + R_l(\infty,x_0;k) e^{-ik(x-x_0)}\quad {\rm for} \quad x<x_0,
\\
\label{1-1.8b}
\phi(x)& \to 0 \quad {\rm as} \quad x \to +\infty.
\end{eqnarray}
\endnumparts
Our objective is to express the Green function in terms of these two quantities, $R_r(x_0,-\infty;k)$ and $R_l(\infty,x_0;k)$. 
The results are applicable to any $V_{\rm S}(x)$ (which is either finite or $+\infty$ at $x=\pm \infty$) as long as the reflection coefficients can be defined for it.

\section{Boson representation}

It was shown in~\cite{structure} that the Green function can be expressed in a general form in terms of the Lie superalgebra $osp(1/2)$. We can obtain various expressions  of the Green function by writing this general expression in specific representations. 
Here we use a representation in terms of boson operators, which is convenient for the methods we shall use in this paper.

Let $a$ and $a^\dagger$ be the boson annihilation-creation operators, satisfying the commutation relation
\begin{equation}
\label{1-2.3}
[a,a^\dagger]=1,
\end{equation}
and let $\vert 0 \rangle$ be the boson vacuum state, satisfying
\begin{equation}
\label{1-2.4}
a\vert 0 \rangle =0, \qquad
\langle 0 \vert a^\dagger=0, \qquad 
\langle 0 \vert 0 \rangle=1.
\end{equation}
We regard the space coordinate $x$ as playing the role of the time, and consider the ^^ ^^ Hamiltonian"
\begin{equation}
\label{1-2.5}
H(x)\equiv
-k \left(a^\dagger a + \case{1}{2}\right)-\case{1}{2}if(x)(aa-a^\dagger a^\dagger),
\end{equation}
where $f(x)$ is the function defined by (\ref{1-1.3}).
The free part of this hamiltonian, $-k a^\dagger a$ describes free propagation of the boson. The interaction part consists of pair creation and pair annihilation of bosons. 
(The constant term $-\case{1}{2}k$ is added for later convenience.)
We define the evolution operator $U(x,x_0)$ as the solution of the differential equation
\begin{equation}
\label{1-2.6}
i \frac{\partial}{\partial x}U(x,x_0)=
H(x)
U(x,x_0)
\end{equation}
with the initial condition $U(x_0,x_0)=1$.
Using this evolution operator, $G(x,x')$ defined by (\ref{1-2.2}) can be written as~\cite{boson}
\begin{equation}
\label{1-2.7}
G(x,x';k)=
\frac
{\langle 0 \vert U(\infty,x)(a + a^\dagger)U(x,x')(a+a^\dagger)U(x',-\infty) \vert 0 \rangle}
{\langle 0 \vert U(\infty,-\infty) \vert 0 \rangle}.
\end{equation}
This is a specific form of the general algebraic expression mentioned above.
To understand the meaning of this expression, it is helpful to think about the expansion of the right-hand side in powers of $f$. This expansion can be visualized by using Feynman diagrams.
Graphically, the right-hand side of (\ref{1-2.7}) is obtained as the sum of all connected diagrams like the one shown in figure~2(a). (Disconnected diagrams are cancelled by the vacuum amplitude in the denominator.) Each diagram represents a path connecting the points $x'$ and $x$. 
The rules for interpreting the diagrams are shown in figure~2(b). 
It should be noted that the expression (\ref{1-2.7}) itself is valid even when the expansion in terms of $f$ is not well defined, e.g., when $f(x)$ is infinite at $x=\pm \infty$. 
\begin{figure}
\hspace{2cm}
\includegraphics[scale=0.7]{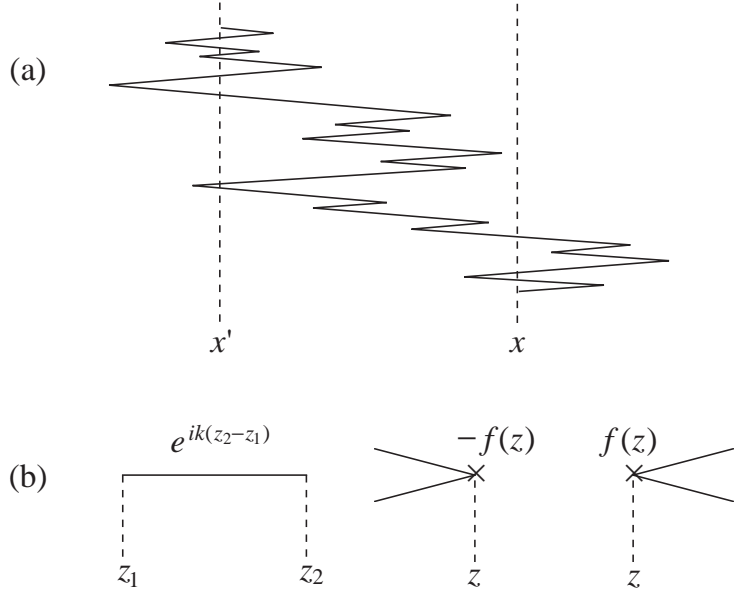}
\caption{
(a) A typical diagram connecting the points $x'$ and $x$. 
(The vertical direction of this figure does not have any particular meaning.)
(b) The diagrammatic rules. Each line segment connecting $x_1$ and $x_2$ corresponds to the free propagator $e^{ik(x_2-x_1)}$. To each turning point of the path is assigned a factor $\pm f(z)$, where the sign is plus if the path comes to that point from the right, and minus if it comes from the left.
}
\end{figure}

\section{Scattering coefficients and the Green function}

The scattering coefficients for a finite interval $(x_1,x_2)$ are defined in the same way as  the reflection coefficients for semi-imfinte intervals we have already introduced.
We consider the Fokker-Planck potential
\refstepcounter{equation}
\label{}
\addtocounter{equation}{-1}
\begin{equation}
\bar V(x)=
\cases{
V(x_1) & $x<x_1,$  \\
V(x) & $x_1 \le x \le x_2,$ \\
V(x_2) & $x_2<x$.
}
\end{equation}
Equation~(\ref{1-1.2}) with $f(x)$ replaced by $\bar  f(x) \equiv -(1/2)(d/dx)\bar V(x)$ has two independent solutions of the form
\numparts
\begin{equation}
\phi_1(x)=
\cases{
e^{[V(x_2)-V(x_1)]/2} \tau(x_2,x_1;k) e^{-ik(x-x_1)} & $x<x_1$, \\
e^{-ik(x-x_2)} + R_r(x_2,x_1;k) e^{ik(x-x_2)} & $x_2<x$, \\
}
\end{equation}
\begin{equation}
\phi_2(x)=
\cases{
e^{ik(x-x_1)} + R_l(x_2,x_1;k) e^{-ik(x-x_1)} &$x<x_1$, \\
e^{-[V(x_2)-V(x_1)]/2} \tau(x_2,x_1;k) e^{ik(x-x_2)} &  $x_2<x$. \\
}
\end{equation}
\endnumparts
This defines the transmission coefficient $\tau$, the right reflection coefficient $R_r$, and the left reflection coefficient $R_l$ for the interval $(x_1,x_2)$. 
In the boson representation, they can be written as~\cite{boson}
\refstepcounter{equation}
\label{1-2.8}
\addtocounter{equation}{-1}
\numparts
\begin{equation}
\label{1-2.8a}
\tau(x_2,x_1)=
\frac{\langle 0 \vert a U(x_2,x_1) a^\dagger \vert 0 \rangle}
{\langle 0 \vert U(x_2,x_1) \vert 0 \rangle},
\end{equation} 
\begin{equation}
\label{1-2.8b}
R_r(x_2,x_1)=
\frac{\langle 0 \vert aa U(x_2,x_1) \vert 0 \rangle}
{\langle 0 \vert U(x_2,x_1) \vert 0 \rangle},
\end{equation}
\begin{equation}
\label{1-2.8c}
R_l(x_2,x_1)=
\frac{\langle 0 \vert U(x_2,x_1) a^\dagger a^\dagger \vert 0 \rangle}
{\langle 0 \vert U(x_2,x_1) \vert 0 \rangle}.
\end{equation}
\endnumparts
The expressions (\ref{1-2.8b}) and (\ref{1-2.8c}) also hold for semi-infinte intervals. Namely, $R_r(x_0,-\infty)$ and $R_l(+\infty,x_0)$ defined by (\ref{1-1.6a}) and (\ref{1-1.8a}) can be expressed as
\begin{equation}
\fl
R_r(x_0,-\infty)=
\frac{\langle 0 \vert aa U(x_0,-\infty) \vert 0 \rangle}
{\langle 0 \vert U(x_0,-\infty) \vert 0 \rangle},
\qquad
R_l(\infty,x_0)=
\frac{\langle 0 \vert U(\infty,x_0) a^\dagger a^\dagger \vert 0 \rangle}
{\langle 0 \vert U(\infty,x_0) \vert 0 \rangle}.
\end{equation}
Similarly to the graphical interpretation of $G(x,x')$ shown in figure~2,
we can interpret (\ref{1-2.8}) in terms of diagrams.
The transmission coefficient $\tau(x_2,x_1)$ is the sum of all the paths that penetrate the interval $(x_1,x_2)$, as in figure~3(a). 
\begin{figure}
\hspace{2cm}
\includegraphics[scale=0.7]{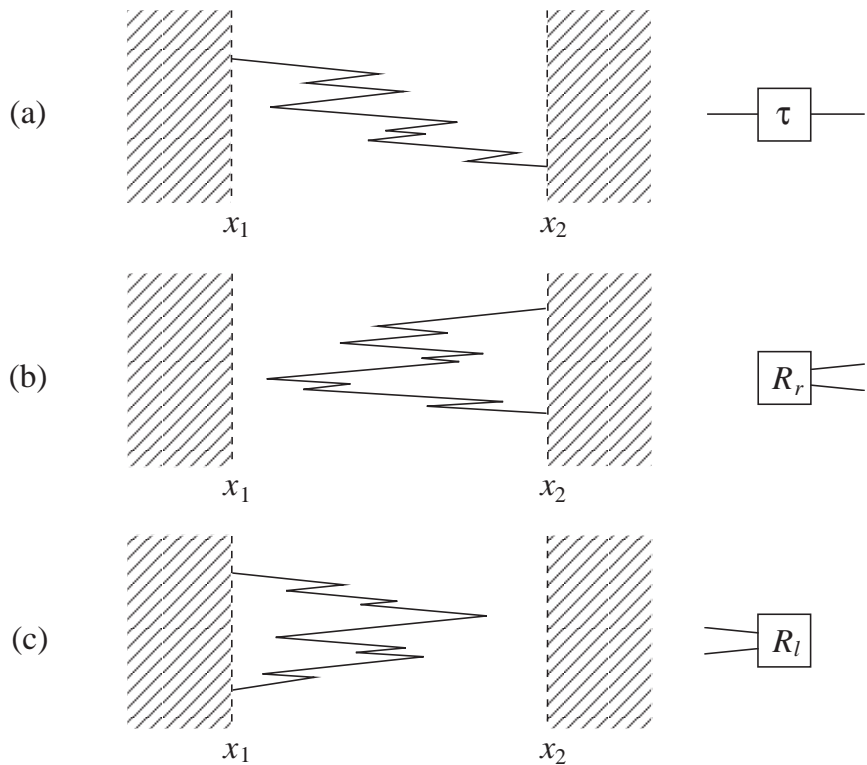}
\caption{
A typical diagram or (a) $\tau$, (b) $R_r$, and (c) $R_l$. 
Such diagrams are to be evaluated with the rules given in figure~2(b).
}
\end{figure}  
The diagrams for the reflection coefficients consist of the paths that start from one of the endpoints of the interval and return to that same point, as shown in figures~3(b) and~3(c).

The scattering coefficients for finite intervals are the quantities that play major roles in the invariant imbedding method. 
We shall use them as building blocks for constructing the full propagator (\ref{1-2.7}).
However, these quantities shall appear only in intermediate steps and not remain in our final results. 
Our objective is to express everything in terms of the reflection coefficients for semi-infinite intervals, without using the quantities (\ref{1-2.8}) for finite intervals. 

As explained in section~2, the propagator $G(x,x')$ is the sum of the paths connecting the points $x'$ and $x$. Such paths can be constructed from the transmission and reflection coefficients.
The idea used here is essentially the same as the old one which dates back to the work by Stokes \cite{stokes}.
\begin{figure}
\hspace{2cm}
\includegraphics[scale=0.7]{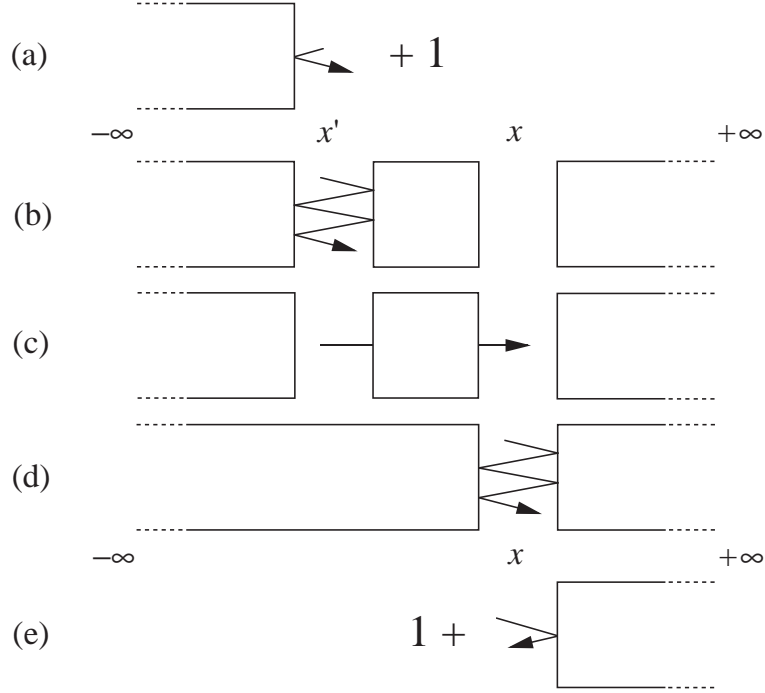}
\caption{
The passage from $x'$ to $x$ can be decomposed into the processes shown here.
They correspond to:
(a) $1+R_r(x',-\infty)$,\, (b) $\sum_{n=0}^\infty [R_l(x,x')R_r(x',-\infty)]^n$,\,
(c) $\tau(x,x')$, \, (d) $\sum_{m=0}^\infty [R_l(\infty,x)R_r(x,-\infty)]^m$, \,
(e) $1+R_l(\infty,x)$.
}
\end{figure}
As illustrated in figure~4, we have~\cite{theory}
\refstepcounter{equation}
\label{1-4.1}
\addtocounter{equation}{-1}
\numparts
\begin{eqnarray}
\label{1-4.1a}
\fl
G(x,x')&= [1+R_l(\infty,x)] 
\biggl(\sum_{n=0}^\infty [R_l(\infty,x)R_r(x,-\infty)]^n \biggr)
\nonumber \\
\fl
& \qquad \qquad \times
\tau(x,x') 
\biggl(\sum_{m=0}^\infty [R_l(x,x')R_r(x',-\infty)]^m \biggr)
 [1+R_r(x',-\infty)].
\end{eqnarray}
Note that this expression is not symmetric with respect to $x$ and $x'$.
This asymmetrical treatment is necessary in order to avoid double counting.
It is also possible to exchange the roles of $x$ and $x'$ in (\ref{1-4.1a}) and write
\begin{eqnarray}
\label{1-4.1b}
\fl
G(x,x')&= [1+R_l(\infty,x)] 
\biggl(\sum_{n=0}^\infty [R_l(\infty,x)R_r(x,x')]^n \biggr)
\nonumber \\
\fl
& \qquad \qquad \times
\tau(x,x')
\biggl(\sum_{m=0}^\infty [R_l(\infty,x')R_r(x',-\infty)]^m \biggr)
[1+R_r(x',-\infty)].
\end{eqnarray}
\endnumparts
The geometric series in equations (\ref{1-4.1}) can be summed to yield
\refstepcounter{equation}
\label{1-4.2}
\addtocounter{equation}{-1}
\numparts
\begin{equation}
\label{1-4.2a}
G(x,x') =\frac{[1+R_l(\infty,x)] [1+R_r(x',-\infty)]\tau(x,x')}
{[1-R_l(\infty,x)R_r(x,-\infty)][1-R_l(x,x')R_r(x',-\infty)]},
\end{equation}
\begin{equation}
\label{1-4.2b}
G(x,x') =\frac{[1+R_l(\infty,x)] [1+R_r(x',-\infty)]\tau(x,x')}
{[1-R_l(\infty,x)R_r(x,x')][1-R_l(\infty,x')R_r(x',-\infty)]}.
\end{equation}
\endnumparts
We wish to eliminate the $\tau(x,x')$, $R_r(x,x')$, and $R_l(x,x')$ from these expressions. 
We shall do this in the next section.

\section{Expressions in terms of reflection coefficients}

The vacuum amplitude $\langle 0 \vert U(x_2,x_1) \vert 0 \rangle$, which appears in the denominators on the right-hand sides of (\ref{1-2.8}), is related to the transmission coefficient by the identity~\cite{structure}
\begin{equation}
\label{1-3.1}
\langle 0 \vert U(x_2,x_1) \vert 0 \rangle =\left[\tau(x_2,x_1)\right]^{1/2}.
\end{equation}
This identity can be checked diagrammatically for each order in $f$ (figure~5).
\begin{figure}
\hspace{2cm}
\includegraphics[scale=0.7]{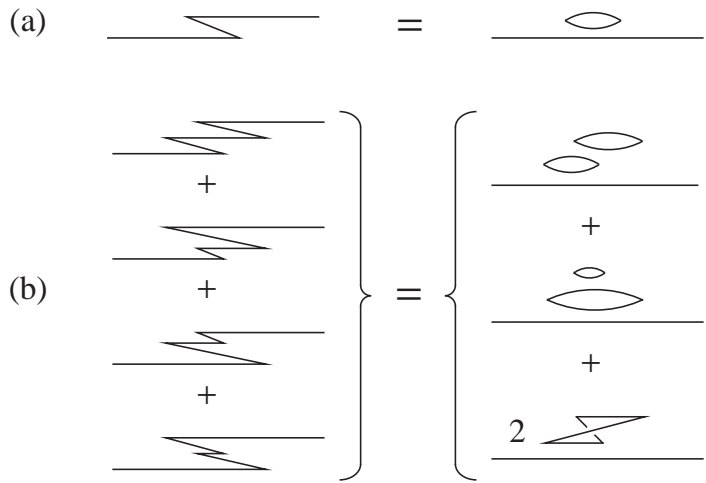}
\caption{
Diagrammatic interpretation of the identity (\ref{1-3.1}) to (a) order $f^2$ and (b) order $f^4$.
On the left-hand sides are the the diagrams of $\tau$. 
On the right-hand sides, the straight lines correspond to $e^{2ik(x_2-x_1)}$, which comes from the constant term $- \case{1}{2}k$ in the Hamiltonian. The bubbles are the diagrams of $Z^2$, where $Z$ is defined by (\ref{1-3.2}). 
(Using the quantity $F$ defined by (\ref{1-3.3}), we can write $Z^2=1+2F+(2F)^2/2+\cdots$. As shown in figure~6, the diagrams for $F$ consist of connected loop  diagrams.)
}
\end{figure}
Since a constant term $- \case{1}{2}k$ is included in the Hamiltonian (\ref{1-2.5}), 
we have $\langle 0 \vert U(x_2,x_1) \vert 0\rangle=e^{ik(x_2-x_1)/2}$ when $f$ is identically zero.
We define
\begin{equation}
\label{1-3.2}
Z(x_2,x_1)\equiv e^{-ik(x_2-x_1)/2}\langle 0 \vert U(x_2,x_1) \vert 0 \rangle,
\end{equation}
so that $Z=1$ when $f=0$. This $Z$ is the vacuum amplitude in the usual sense; the Feynman diagrams for $Z$ are bubble diagrams without external legs. These bubble diagrams are, in general, disconnected. 
To deal with connected Feynman diagrams, we define
\begin{equation}
\label{1-3.3}
F(x_2,x_1) \equiv \log Z(x_2,x_1).
\end{equation}
As is known in usual diagrammatic discussions in field theory~\cite{zinnjustin}, this $F$ is obtained as the sum of all connected loop diagrams (figure~6(a)).
(In statistical mechanics, $Z$ and $F$ correspond to the partition function and the free energy, respectively.)
\begin{figure}
\hspace{2cm}
\includegraphics[scale=0.7]{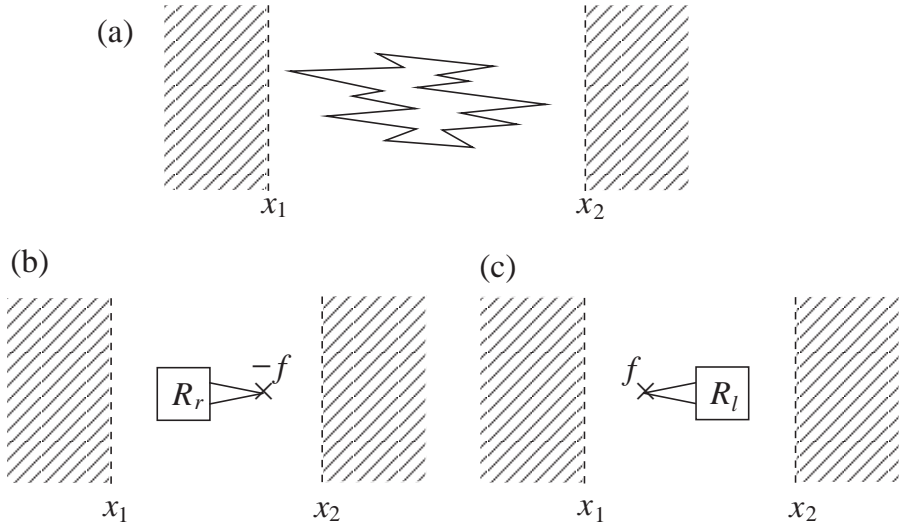}
\caption{
(a) A typical diagram of $F(x_2,x_1)$. 
As shown in (b) and (c), such a diagram can be obtained from a diagram for $R_r$ or $R_l$ (figures~3(b) and (c)).
}
\end{figure}
From (\ref{1-3.1}), (\ref{1-3.2}), and (\ref{1-3.3}), we have
\begin{equation}
\label{1-3.4}
\tau(x_2,x_1)=\exp\left[ik(x_2-x_1)+2F(x_2,x_1)\right].
\end{equation}

Connected loop diagrams are obtained by connecting the two legs of $R_r$ with a factor $-f$ (see figure~6(b)). This fact can be expressed as
\refstepcounter{equation}
\label{1-3.5}
\addtocounter{equation}{-1}
\numparts
\begin{equation}
\label{1-3.5a}
F(x_2,x_1)=-\frac{1}{2}\int_{x_1}^{x_2}f(z) R_r(z,x_1)\,dz.
\end{equation}
(There is a factor $1/2$ because the same diagram is obtained by exchanging the two legs of $R_r$.)
In the same way, $F$ can also be obtained from $R_l$ as shown in figure~6(c). So we have
\begin{equation}
\label{1-3.5b}
F(x_2,x_1)=\frac{1}{2}\int_{x_1}^{x_2}f(z) R_l(x_2,z)\,dz.
\end{equation}
\endnumparts
In the invariant imbedding method, differential equations satisfied by the scattering coefficients are derived by varying an endpoint of the interval. 
Analogous differential equations for the quantity $F$ are obtained from equations (\ref{1-3.5}) as
\begin{equation}
\label{1-3.7}
\fl
\frac{\partial}{\partial x_2} F(x_2,x_1)=-\frac{1}{2}f(x_2)R_r(x_2,x_1),
\qquad 
\frac{\partial}{\partial x_1} F(x_2,x_1)=-\frac{1}{2}f(x_1)R_l(x_2,x_1).
\end{equation}
There is another useful relation that connects $F$ to the reflection coefficients:
\begin{equation}
\label{1-3.6}
\fl
F(x_c,x_a)-F(x_c,x_b)-F(x_b,x_a)
=-\frac{1}{2}\log\left[1-R_l(x_c,x_b)R_r(x_b,x_a)\right],
\end{equation}
where $x_a \leq x_b \leq x_c$.
We can understand this relation diagrammatically. 
The left-hand side of (\ref{1-3.6}) is the sum of all connected loop diagrams which are restricted within the interval $(x_a,x_c)$, and which cross the point $x_b$ (figure~7(a)).
\begin{figure}
\hspace{2cm}
\includegraphics[scale=0.7]{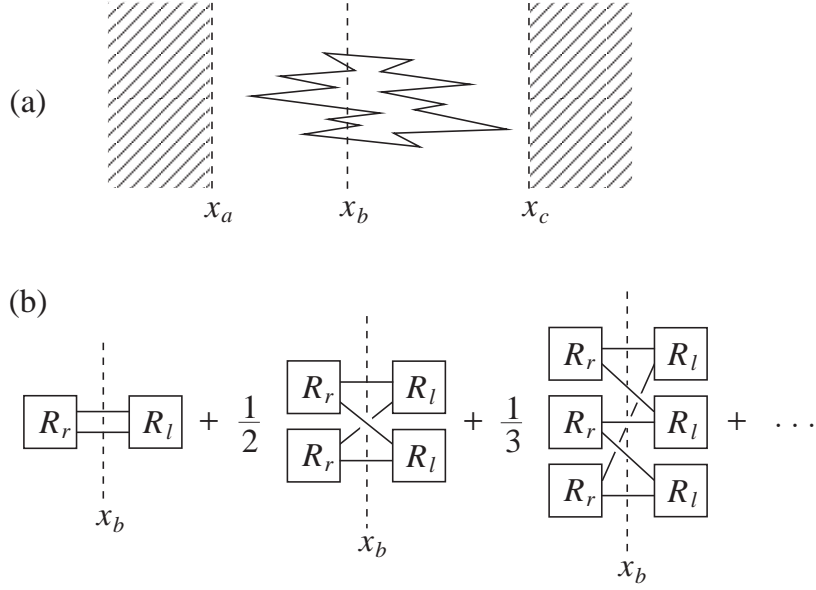}
\caption{
(a) A diagram contributing to $F(x_c,x_a)-F(x_c,x_b)-F(x_b,x_a)$.
(b) Such diagrams can be constructed in this way, using the reflection coefficients.
(The factors $\case{1}{2}$, $\case{1}{3}$, etc are necessary in order to avoid double counting.) 
}
\end{figure}
Such diagrams are obtained from the reflection coefficients as shown in figure~7(b).  
The series in figure~7(b) can be summed as
\begin{equation}
R_lR_r+\case{1}{2}(R_lR_r)^2+\case{1}{3}(R_lR_r)^3+\cdots=-\log(1-R_lR_r).
\end{equation}
Hence we have (\ref{1-3.6}). (There is an overall factor $\case{1}{2}$ on the right-hand side of (\ref{1-3.6}) for the same reason as in equations (\ref{1-3.5}).)
All the relations such as (\ref{1-3.5}) or (\ref{1-3.6}) are valid even when the expansion in terms of $f$ is not well defined. (It is not difficult to prove these relations without using the diagrams.)

Differentiating both sides of (\ref{1-3.6}) with respect to $x_a$, $x_b$, or $x_c$, and using 
(\ref{1-3.7}), we obtain
\refstepcounter{equation}
\label{1-3.8}
\addtocounter{equation}{-1}
\numparts
\begin{equation}
\label{1-3.8a}
\fl
\frac{\partial}{\partial x_c}
\log\left[1-R_l(x_c,x_b)R_r(x_b,x_a)\right]
=f(x_c)\left[R_r(x_c,x_a)-R_r(x_c,x_b)\right],
\end{equation}
\begin{equation}
\label{1-3.8b}
\fl
\frac{\partial}{\partial x_b}
\log\left[1-R_l(x_c,x_b)R_r(x_b,x_a)\right]
=-f(x_b)\left[R_l(x_c,x_b)+R_r(x_b,x_a)\right],
\end{equation}
\begin{equation}
\label{1-3.8c}
\fl
\frac{\partial}{\partial x_a}
\log\left[1-R_l(x_c,x_b)R_r(x_b,x_a)\right]
=f(x_a)\left[R_l(x_c,x_a)-R_l(x_b,x_a)\right].
\end{equation}
\endnumparts
Setting $x_c=z$, $x_b=x_1$, $x_a=-\infty$ in (\ref{1-3.8a}), and integrating both sides with respect to $z$ from $x_1$ to $x_2$, we have
\begin{equation}
\label{1-3.9}
\fl
\log \left[1-R_l(x_2,x_1)R_r(x_1,-\infty)\right]
=\int_{x_1}^{x_2} f(z) \left[R_r(z,-\infty)-R_r(z,x_1)\right]\,dz,
\end{equation}
where we have used $R_r(x_1,x_1)=0$. 
Using (\ref{1-3.9}), we can rewrite (\ref{1-3.5a}) as
\begin{equation}
\label{1-3.10}
\fl
F(x_2,x_1)=-\frac{1}{2}\int_{x_1}^{x_2}f(z) R_r(z,-\infty)\,dz
+\frac{1}{2}\log \left[1-R_l(x_2,x_1)R_r(x_1,-\infty)\right].
\end{equation}
Substituting this into (\ref{1-3.4}) yields
\refstepcounter{equation}
\label{1-3.11}
\addtocounter{equation}{-1}
\numparts
\begin{equation}
\label{1-3.11a}
\fl
\tau(x,x')=
\left[1-R_l(x,x')R_r(x',-\infty)\right]
\exp\left[ik(x-x')-\int_{x'}^x f(z) R_r(z,-\infty)\,dz \right].
\end{equation}
In the same way, using (\ref{1-3.5b}) and (\ref{1-3.8c}) we can derive
\begin{equation}
\label{1-3.11b}
\fl
\tau(x,x')=
\left[1-R_l(\infty,x)R_r(x,x')\right]
\exp\left[ik(x-x')+\int_{x'}^x f(z) R_l(\infty,z)\,dz \right].
\end{equation}
\endnumparts

On substituting (\ref{1-3.11a}) into (\ref{1-4.2a}), the factor including $R_l(x,x')$ cancels out,
and $G(x,x')$ is expressed solely in terms of reflection coefficients for semi-infinite intervals:
\refstepcounter{equation}
\label{1-4.3}
\addtocounter{equation}{-1}
\numparts
\begin{eqnarray}
\label{1-4.3a}
\fl
G(x,x') &=
\frac{[1+R_l(\infty,x)] [1+R_r(x',-\infty)]}{1-R_l(\infty,x)R_r(x,-\infty)}
\exp\left[ik(x-x')-\int_{x'}^x f(z) R_r(z,-\infty)\,dz \right].
\nonumber \\
\fl
\end{eqnarray}
Similarly, from (\ref{1-3.11b}) and (\ref{1-4.2b}) we obtain
\begin{eqnarray}
\label{1-4.3b}
\fl
G(x,x') &=
\frac{[1+R_l(\infty,x)] [1+R_r(x',-\infty)]}{1-R_l(\infty,x')R_r(x',-\infty)}
\exp\left[ik(x-x')+\int_{x'}^x f(z) R_l(\infty,z)\,dz \right].
\nonumber \\
\fl
\end{eqnarray}
\endnumparts
On the other hand, integrating (\ref{1-3.8b}) and setting $x_c=\infty$, $x_a=-\infty$ gives
\begin{equation}
\label{1-4.4}
\fl
\int_{x'}^x f(z) [R_l(\infty,z)+R_r(z,-\infty)]\,dz
=
\log\frac{1-R_l(\infty,x')R_r(x',-\infty)}{1-R_l(\infty,x)R_r(x,-\infty)}.
\end{equation}
It is obvious that (\ref{1-4.3a}), (\ref{1-4.3b}), and (\ref{1-4.4}) are consistent.
A symmetric expression of $G(x,x')$ is obtained by multiplying (\ref{1-4.3a}) and (\ref{1-4.3b}), and taking the square root:
\begin{eqnarray}
\label{1-4.5}
\fl
G(x,x')&=\frac{[1+R_l(\infty,x)][1+R_r(x',-\infty)] e^{ik(x-x')}}
{[1-R_l(\infty,x)R_r(x,-\infty)]^{1/2}[1-R_l(\infty,x')R_r(x',-\infty)]^{1/2} }
\nonumber \\
\fl
&\qquad \qquad \times \exp\left\{
\frac{1}{2} \int_{x'}^x f(z)[R_l(\infty,z)-R_r(z,-\infty)]\,dz
\right\}.
\end{eqnarray}

\section{Generalization}
In this section we shall derive a more general expression of the Green function, which includes (\ref{1-4.5}) as a special case.
The derivation is based on a symmetry transformation which can be understood as a rotation of the coordinate axes~\cite{algebraic}. 

We define
\begin{equation}
\label{1-5.1}
X(x) \equiv ikx, \qquad Y(x) \equiv V(x)/2,
\end{equation}
where $V(x)$ is the Fokker-Planck potential. 
Then (\ref{1-2.6}) with (\ref{1-2.5}) can be written as
\begin{equation}
\label{1-5.2}
\frac{\partial}{\partial x}U(x,x_0)=
\frac{1}{2}
\left[
\frac{dX}{dx} (a a^\dagger+a^\dagger a)+\frac{dY}{dx}(aa-a^\dagger a^\dagger)
\right]
U(x,x_0).
\end{equation}
We consider the rotation of the $X$-$Y$ axes, defining
\begin{equation}
\label{1-5.3}
\left(
\begin{array}{cc}
X_\theta \\
Y_\theta \\
\end{array}
\right)
\equiv
\left(
\begin{array}{cc}
\cos \theta & \sin \theta \\
-\sin \theta &\cos \theta \\
\end{array}
\right)
\left(
\begin{array}{cc}
X \\
Y \\
\end{array}
\right).
\end{equation}
We also define the boson operators in the rotated frame as
\begin{equation}
\label{1-5.4}
\left(
\begin{array}{cc}
a_\theta \\
a^\dagger_\theta \\
\end{array}
\right)
\equiv
\left(
\begin{array}{cc}
\cos \frac{\theta}{2} & -\sin \frac{\theta}{2} \\
\sin \frac{\theta}{2} & \cos \frac{\theta}{2} \\
\end{array}
\right)
\left(
\begin{array}{cc}
a \\
a^\dagger \\
\end{array}
\right).
\end{equation}
They are indeed boson operators, satisfying the commutation relation
\begin{equation}
\label{1-5.5}
[a_\theta,a_\theta^\dagger]=1.
\end{equation}
It is easy to see that equation (\ref{1-5.2}) is covariant under this rotation;
it sill holds when $X$, $Y$, $a$, $a^\dagger$ are replaced by the ones with subscript $\theta$:
\begin{equation}
\label{1-5.6}
\frac{\partial}{\partial x}U(x,x_0)=
\frac{1}{2}
\left[
\frac{dX_\theta}{dx} (a_\theta a_\theta^\dagger+a_\theta^\dagger a_\theta)
+\frac{dY_\theta}{dx}(a_\theta a_\theta-a_\theta^\dagger a_\theta^\dagger)
\right]
U(x,x_0).
\end{equation}
Let $\vert 0;\theta \rangle$ denote the vacuum state in the rotated frame, satisfying
\begin{equation}
\label{1-5.7}
a_\theta \vert 0;\theta \rangle=0, \qquad
\langle 0;\theta \vert a_\theta^\dagger=0, \qquad 
\langle 0;\theta \vert 0;\theta \rangle=1.
\end{equation}
It can be shown that this state is related to the original vacuum as~\cite{boson}
\refstepcounter{equation}
\label{1-5.8}
\addtocounter{equation}{-1}
\numparts
\begin{equation}
\label{1-5.8a}
\vert 0;\theta \rangle=(1+\eta^2)^{1/4} 
\exp(\eta \,a^\dagger a^\dagger/2)
\vert 0 \rangle,
\end{equation}
\begin{equation}
\label{1-5.8b}
\langle 0;\theta \vert=(1+\eta^2)^{1/4}  
\langle 0 \vert
\exp(-\eta \,a a/2),
\end{equation}
\endnumparts
\refstepcounter{equation}
\label{1-5.9}
\addtocounter{equation}{-1}
\numparts
\begin{equation}
\label{1-5.9a}
\vert 0 \rangle=(1+\eta^2)^{1/4} 
\exp(-\eta \,a_\theta^\dagger a_\theta^\dagger/2)
\vert 0;\theta \rangle,
\end{equation}
\begin{equation}
\label{1-5.9b}
\langle 0 \vert=(1+\eta^2)^{1/4}  
\langle 0;\theta \vert
\exp(\eta \,a_\theta a_\theta/2),
\end{equation}
\endnumparts
where
\begin{equation}
\label{1-5.10}
\eta \equiv \tan \case{\theta}{2}.
\end{equation}
Using (\ref{1-5.9}), we can rewrite (\ref{1-2.7}) as
\begin{eqnarray}
\label{1-5.11}
\fl
G(x,x')=
\frac
{\langle 0;\theta \vert
e^{\eta a_\theta a_\theta/2}
U(\infty,x)(a + a^\dagger)U(x,x')(a+a^\dagger)U(x',-\infty)
e^{-\eta a_\theta^\dagger a_\theta^\dagger/2}
\vert 0;\theta \rangle}
{\langle 0;\theta \vert
e^{\eta a_\theta a_\theta/2}
 U(\infty,-\infty) 
e^{-\eta a_\theta^\dagger a_\theta^\dagger/2}
\vert 0;\theta \rangle},
\nonumber \\
\fl
\end{eqnarray}
where
\begin{equation}
\label{1-5.12}
a+a^\dagger
=\left(\cos \case{\theta}{2}-\sin \case{\theta}{2}\right)a_\theta
+\left(\cos \case{\theta}{2}+\sin \case{\theta}{2}\right)a_\theta^\dagger.
\end{equation}
Equation (\ref{1-5.11}) holds for any $\theta$, and so it is a generalized form of (\ref{1-2.7}).

Just like (\ref{1-2.8}), we define the scattering coefficients in the rotated frame:
\refstepcounter{equation}
\label{1-5.13}
\addtocounter{equation}{-1}
\numparts
\begin{equation}
\label{1-5.13a}
\tau_\theta(x_2,x_1)\equiv 
\frac{\langle 0 ;\theta \vert a_\theta U(x_2,x_1) a_\theta^\dagger \vert 0;\theta \rangle}
{\langle 0;\theta \vert U(x_2,x_1) \vert 0;\theta \rangle},
\end{equation}
\begin{equation}
\label{1-5.13b}
R_{r,\theta}(x_2,x_1)
\equiv \frac{\langle 0;\theta \vert a_\theta a_\theta U(x_2,x_1) \vert 0;\theta \rangle}
{\langle 0;\theta \vert U(x_2,x_1) \vert 0;\theta \rangle},
\end{equation}
\begin{equation}
\label{1-5.13c}
R_{l,\theta}(x_2,x_1)
\equiv \frac{\langle 0;\theta \vert U(x_2,x_1) a_\theta^\dagger a_\theta^\dagger \vert 0;\theta \rangle}
{\langle 0;\theta \vert U(x_2,x_1) 
\vert 0;\theta \rangle}.
\end{equation}
\endnumparts
Since the evolution equation (\ref{1-5.6}) has the same form as (\ref{1-5.2}), 
these scattering coefficients can be interpreted diagrammatically in the same way as before (i.e., as in figure~3).
The rules in figure~2(b) are now generalized to the ones shown in figure~8;
\begin{figure}
\hspace{2cm}
\includegraphics[scale=0.7]{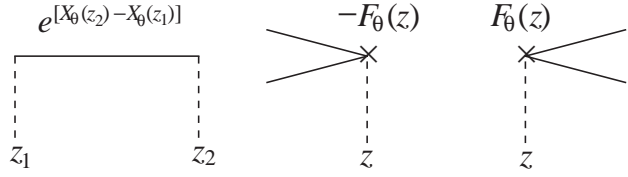}
\caption{
The diagrammatic rules in the rotated frame with angle $\theta$.
}
\end{figure}
as can be seen from the right-hand side of (\ref{1-5.6}), the free propagator connecting $x_1$ and $x_2$ is now $\exp[X_\theta(x_2)-X_\theta(x_1)]$, and the factor assigned to each turning point $z$ is $\pm F_\theta(z)$, where
\begin{equation}
\label{1-5.14}
F_\theta(x)\equiv -\frac{dY_\theta(x)}{dx}.
\end{equation}
Since the expression (5.11) involves the state $e^{-\eta a_\theta^\dagger a_\theta^\dagger/2}
\vert 0;\theta \rangle$, it is convenient to define, in addition to (\ref{1-5.13}), the quantities
\refstepcounter{equation}
\label{5.15}
\addtocounter{equation}{-1}
\numparts
\begin{equation}
\label{1-5.15a}
\rho_{r,\theta}(x)
\equiv \frac{\langle 0;\theta \vert 
a_\theta a_\theta U(x,-\infty) 
e^{-\eta a_\theta^\dagger a_\theta^\dagger/2}
\vert 0;\theta \rangle}
{\langle 0;\theta \vert U(x,-\infty) 
e^{-\eta a_\theta^\dagger a_\theta^\dagger/2}
\vert 0;\theta \rangle},
\end{equation}
\begin{equation}
\label{1-5.15b}
\rho_{l,\theta}(x)
\equiv \frac{\langle 0;\theta \vert 
e^{\eta a_\theta a_\theta/2}
U(\infty,x) a_\theta^\dagger a_\theta^\dagger \vert 0;\theta \rangle}
{\langle 0;\theta \vert
e^{\eta a_\theta a_\theta/2}
 U(\infty,x) \vert 0;\theta \rangle}.
\end{equation}
\endnumparts
They can be interpreted as reflection coefficients including additional scattering at infinity (see figure~9).
\begin{figure}
\hspace{2cm}
\includegraphics[scale=0.7]{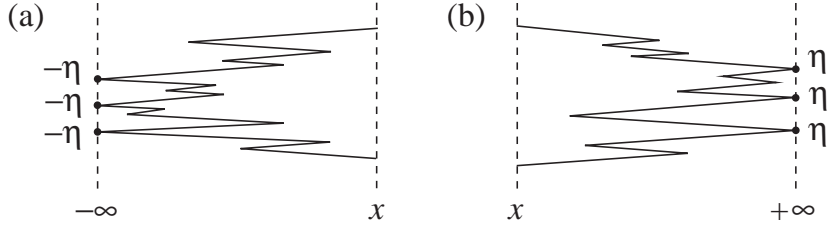}
\caption{
Diagrammatic interpretation of (a) $\rho_{r,\theta}(x)$, and (b) $\rho_{l,\theta}(x)$.
Now the path is reflected at infinity. Each reflection at $\pm \infty$ gives a factor $\pm \eta$.  
}
\end{figure}

The expressions in the rotated frame corresponding to (\ref{1-3.11a}) and (\ref{1-3.11b}) are obtained by adding the subscript $\theta$ to the scattering coefficients, and making the replacements $ikx \to X_\theta(x)$ and $f(z)\to F_\theta(z)$: 
\refstepcounter{equation}
\label{5.22}
\addtocounter{equation}{-1}
\numparts
\begin{eqnarray}
\label{1-5.22a}
\fl
\tau_\theta(x,x')=
\left[1-R_{l,\theta}(x,x')R_{r,\theta}(x',-\infty)\right]
\exp\left[X_\theta(x)-X_\theta(x')-\int_{x'}^x F_\theta(z) R_{r,\theta}(z,-\infty)\,dz \right],
\nonumber \\
\fl
\end{eqnarray}
\begin{eqnarray}
\label{1-5.22b}
\fl
\tau_\theta(x,x')=
\left[1-R_{l,\theta}(\infty,x)R_{r,\theta}(x,x')\right]
\exp\left[X_\theta(x)-X_\theta(x')+\int_{x'}^x F_\theta(z) R_{l,\theta}(\infty,z)\,dz \right].
\nonumber \\
\fl
\end{eqnarray}
\endnumparts
From the derivation of (\ref{1-3.11}), and from the diagrammatic interpretation of the quantities $\rho_{r,\theta}$ and $\rho_{l,\theta}$ shown in figure~9, it is obvious that these expressions still hold when $R_{r,\theta}(z,-\infty)$ and $R_{l,\theta}(\infty,z)$ are replaced by $\rho_{r,\theta}(z)$ and $\rho_{l,\theta}(z)$, respectively:
\refstepcounter{equation}
\label{1-5.23}
\addtocounter{equation}{-1}
\numparts
\begin{equation}
\label{1-5.23a}
\fl
\tau_\theta(x,x')=
\left[1-R_{l,\theta}(x,x')\rho_{r,\theta}(x')\right]
\exp\left[X_\theta(x)-X_\theta(x')-\int_{x'}^x F_\theta(z) \rho_{r,\theta}(z)\,dz \right],
\nonumber \\
\fl
\end{equation}
\begin{equation}
\label{1-5.23b}
\fl
\tau_\theta(x,x')=
\left[1-\rho_{l,\theta}(x)R_{r,\theta}(x,x')\right]
\exp\left[X_\theta(x)-X_\theta(x')+\int_{x'}^x F_\theta(z) \rho_{l,\theta}(z)\,dz \right].
\nonumber \\
\fl
\end{equation}
\endnumparts

Comparing (\ref{1-2.7}) with (\ref{1-5.11}), we find that the generalized forms of (\ref{1-4.2a}) and (\ref{1-4.2b}) are obtained by making the following replacements:
First, the scattering coefficients for the interval $(x',x)$ are replaced by the quantities with subscript $\theta$:
\begin{equation}
\label{1-5.24}
\fl
\tau(x,x') \to \tau_\theta(x,x'), \qquad
R_r(x,x') \to R_{r,\theta}(x,x'), \qquad
R_l(x,x') \to R_{l,\theta}(x,x').
\end{equation}
Second, the reflection coefficients for semi-infinite intervals are replaced not by $R_{r,\theta}$ and $R_{l,\theta}$ but by $\rho_{r, \theta}$ and $\rho_{l,\theta}$:
\begin{equation}
\label{1-5.25}
R_r(z,-\infty) \to \rho_{r,\theta}(z), 
\qquad
R_l(\infty,z) \to \rho_{l,\theta}(z),
\end{equation}
where $z$ is either $x$ or $x'$.
This is because the state $e^{-\eta a_\theta^\dagger a_\theta^\dagger/2}
\vert 0;\theta \rangle$ appears in (\ref{1-5.11}) instead of $\vert 0;\theta \rangle$. 
Third, since the operator $a+a^\dagger$ is replaced by the right-hand side of (\ref{1-5.12}),
the expression $[1+R_l(\infty,x)][1+R_r(x,-\infty)]$ in (\ref{1-4.2}) needs to be replaced as
\begin{equation}
\label{1-5.26}
[1+R_l(\infty,x)] [1+R_r(x',-\infty)] \to A_\theta(x,x'),
\end{equation}
where
\begin{eqnarray}
\label{1-5.27}
A_\theta(x,x')
&\equiv 
\left[
\cos \case{\theta}{2} -\sin \case{\theta}{2}
+ \left(\cos \case{\theta}{2} + \sin \case{\theta}{2} \right)\rho_{l,\theta}(x)
\right]
\nonumber \\
& \qquad \times
\left[
\cos \case{\theta}{2} +\sin \case{\theta}{2}
+ \left(\cos \case{\theta}{2} - \sin \case{\theta}{2} \right)\rho_{r,\theta}(x')
\right].
\end{eqnarray}
Making these replacements in (\ref{1-4.2}) leads to
\refstepcounter{equation}
\label{1-5.28}
\addtocounter{equation}{-1}
\numparts
\begin{equation}
\label{1-5.28a}
G(x,x')=
\frac
{
A_\theta(x,x')\,
\tau_\theta(x,x')
}
{
\left[1-\rho_{l,\theta}(x)\rho_{r,\theta}(x)\right]
\left[1-R_{l,\theta}(x,x')\rho_{r,\theta}(x')\right]
},
\end{equation}
\begin{equation}
\label{1-5.28b}
G(x,x')=
\frac
{
A_\theta(x,x')\,
\tau_\theta(x,x')
}
{
\left[1-\rho_{l,\theta}(x)R_{r,\theta}(x,x')\right]
\left[1-\rho_{l,\theta}(x')\rho_{r,\theta}(x')\right]
}.
\end{equation}
\endnumparts
Substituting (\ref{1-5.23a}) and (\ref{1-5.23b}) into (\ref{1-5.28a}) and (\ref{1-5.28b}), respectively, we have
\refstepcounter{equation}
\label{1-5.29}
\addtocounter{equation}{-1}
\numparts
\begin{equation}
\label{1-5.29a}
\fl
G(x,x')=
\frac{A_\theta(x,x')}{1-\rho_{l,\theta}(x)\rho_{r,\theta}(x)}
\exp
\left[
X_\theta(x)-X_\theta(x')-\int_{x'}^x F_\theta(z) \rho_{r,\theta}(z)\,dz
\right],
\end{equation}
\begin{equation}
\label{1-5.29b}
\fl
G(x,x')=
\frac{A_\theta(x,x')}{1-\rho_{l,\theta}(x')\rho_{r,\theta}(x')}
\exp
\left[
X_\theta(x)-X_\theta(x')+\int_{x'}^x F_\theta(z) \rho_{l,\theta}(z)\,dz
\right],
\end{equation}
\endnumparts
where
\begin{equation}
\label{1-5.30}
\fl
X_\theta(x) = ikx \cos \theta + \case{1}{2}V(x) \sin \theta,
\qquad 
F_\theta(x) = ik \sin \theta + f(x) \cos \theta.
\end{equation}
Since equations (\ref{1-5.29}) hold for any $\theta$, we may let $\theta\to -\theta$ in (\ref{1-5.29b}):
\begin{equation}
\label{1-5.31}
\fl
G(x,x')=
\frac{A_{-\theta}(x,x')}
{1-\rho_{l,-\theta}(x')\rho_{r,-\theta}(x')}
\exp
\left[
X_{-\theta}(x)-X_{-\theta}(x')+\int_{x'}^x F_{-\theta}(z) \rho_{l,-\theta}(z)\,dz
\right].
\end{equation}
A symmetric expression is obtained by taking the geometric mean of (\ref{1-5.29a}) and (\ref{1-5.31}):
\begin{eqnarray}
\label{2-5.26}
\fl
G(x,x')&=
\left(
\frac{A_\theta(x,x')A_{-\theta}(x,x')}
{[1-\rho_{l,\theta}(x)\rho_{r,\theta}(x)][1-\rho_{l,-\theta}(x')\rho_{r,-\theta}(x')]}
\right)^{1/2}
\nonumber \\
\fl
& \qquad \times
\exp
\left\{
\frac{1}{2}\left[
X_\theta(x)+X_{-\theta}(x)-X_\theta(x')-X_{-\theta}(x')\right]
\right\}
\nonumber \\
\fl
& \qquad \times
\exp
\left\{
-\frac{1}{2}\int_{x'}^x 
\left[
F_\theta(z) \rho_{r,\theta}(z)-F_{-\theta}(z) \rho_{l,-\theta}(z)
\right]dz
\right\}.
\end{eqnarray}
Thus, we have obtained expressions for $G(x,x')$ in terms of $\rho_{r,\theta}$ and $\rho_{l,\theta}$.

The quantities $\rho_{r,\theta}$ and $\rho_{l,\theta}$ can be expressed in terms of $R_r(x,-\infty)$ and $R_l(\infty,x)$, as we shall now see.
Using (\ref{1-5.8}), (\ref{1-5.9}) and (\ref{1-5.4}), we write (\ref{1-5.15a}) as
\begin{equation}
\label{1-5.16}
\fl
\rho_{r,\theta}(x)
=
\frac{\langle 0 \vert
e^{-\eta a a/2}
(\cos \case{\theta}{2} a -\sin \case{\theta}{2} a^\dagger)
(\cos \case{\theta}{2} a -\sin \case{\theta}{2} a^\dagger) 
U(x,-\infty) 
\vert 0\rangle}
{\langle 0\vert 
e^{-\eta a a/2}
U(x,-\infty) 
\vert 0\rangle}.
\end{equation}
Using the commutation relations
\begin{equation}
\label{1-5.17}
[a^\dagger a,e^{-\eta aa/2}]
=\eta aae^{-\eta aa/2}
\end{equation}
and
\begin{equation}
\label{1-5.18}
[a^\dagger a^\dagger,e^{-\eta aa/2}]
=(\eta + 2\eta a^\dagger a-\eta^2 aa)e^{-\eta aa/2},
\end{equation}
equation (\ref{1-5.16}) can be modified to the form
\begin{equation}
\label{1-5.19}
\rho_{r,\theta}(x)
=-\eta
+(1+\eta^2)
\frac
{{\langle 0\vert 
e^{-\eta a a/2} aa
U(x,-\infty) 
\vert 0\rangle}}
{{\langle 0\vert 
e^{-\eta a a/2}
U(x,-\infty) 
\vert 0\rangle}},
\end{equation}
where we have also used the definition (\ref{1-5.10}).
Using the diagrammatic interpretation, we can easily see that
\begin{equation}
\frac
{{\langle 0\vert 
a^{2n}
U
\vert 0\rangle}}
{{\langle 0\vert 
U
\vert 0\rangle}}
=\frac{(2n)!}{2^n n!}R_r^n.
\end{equation}
(See \cite{structure} for an explanation.) Therefore,
\begin{equation}
\frac
{{\langle 0\vert 
e^{-\eta a a/2}
U
\vert 0\rangle}}
{{\langle 0\vert 
U
\vert 0\rangle}}
=\sum_{n=0}^\infty
\frac{(2n)!}{(2^n n!)^2}(-\eta R_r)^n
=(1+\eta R_r)^{-1/2}.
\end{equation}
Hence we obtain
\begin{equation}
\fl
\frac
{{\langle 0\vert 
e^{-\eta a a/2} aa
U
\vert 0\rangle}}
{{\langle 0\vert 
e^{-\eta a a/2}
U
\vert 0\rangle}}
=-2 \frac{d}{d\eta} 
\log \frac
{{\langle 0\vert 
e^{-\eta a a/2}
U
\vert 0\rangle}}
{{\langle 0\vert 
U
\vert 0\rangle}}
=\frac{d}{d\eta}\log (1+\eta R_r)
=\frac{R_r}{1+\eta R_r}.
\end{equation}
Substituting this into (\ref{1-5.19}) gives
\refstepcounter{equation}
\label{1-5.21}
\addtocounter{equation}{-1}
\numparts
\begin{equation}
\label{1-5.21a}
\rho_{r,\theta}(x)=\frac{R_r(x,-\infty)-\eta}{1+\eta R_r(x,-\infty)}.
\end{equation}
In the same way, we have
\begin{equation}
\label{1-5.21b}
\rho_{l,\theta}(x)=\frac{R_l(\infty,x)+\eta}{1-\eta R_l(\infty,x)}.
\end{equation}
\endnumparts

Substituting (\ref{1-5.30}) and (\ref{1-5.21}) into (\ref{2-5.26}) we obtain, after some calculation,
\begin{eqnarray}
\label{1-5.32}
\fl
G(x,x')&=
\left(
\frac{[1+\eta R_r(x)][1+\eta R_l(x')]}
{[1+\eta R_l(x)][1+\eta R_r(x')]}
\right)^{1/2}
\frac{[1+R_l(x)][1+R_r(x')] e^{ik(x-x')}}
{[1-R_r(x)R_l(x)]^{1/2}[1-R_r(x')R_l(x')]^{1/2}}
\nonumber \\
\fl
&\qquad \times
\exp
\left[
-\eta ik\int_{x'}^x
\left(
\frac{R_r(z)}{1+\eta R_r(z)}+\frac{R_l(z)}{1+\eta R_l(z)}
\right)\,dz
\right]
\nonumber \\
\fl
&\qquad \times
\exp
\left[
-\frac{1}{2}(1-\eta^2)\int_{x'}^x
f(z)
\left(
\frac{R_r(z)}{1+\eta R_r(z)}-\frac{R_l(z)}{1+\eta R_l(z)}
\right)\,dz
\right],
\end{eqnarray}
where $R_r(x)$ and $R_l(x)$ stand for $R_r(x,-\infty)$ and $R_l(\infty,x)$, respectively.
This is the generalized form of (\ref{1-4.5}). 
Equation (\ref{1-5.32}) holds for any $\theta$, i.e., for any real number $\eta=\tan \theta$.
We recover (\ref{1-4.5}) by setting $\theta=0$ ($\eta=0$).

\section{Expression with $\theta=\pi/2$}

With $\theta=\pm \pi/2$ ($\eta=\pm 1$), the right-hand side of (\ref{1-5.32}) takes a form that does not include the function $f(x)$ explicitly.
In particular, we have a very simple expression with $\theta=+\pi/2$.
Let us define
\begin{equation}
\label{1-6.1}
S(x;k) \equiv \frac{R_l(\infty,x;k)}{1+R_l(\infty,x;k)}+\frac{R_r(x,-\infty;k)}{1+R_r(x,-\infty;k)}.
\end{equation}
Setting $\eta=1$ in (\ref{1-5.32}), and using (\ref{1-6.1}), we can write
\begin{equation}
\label{1-6.2}
\fl
G(x,x';k)=
\frac{1}{\sqrt{[1-S(x;k)][1-S(x';k)]}}
\exp\left[
ik(x-x')-ik\int_{x'}^x S(z;k)\,dz
\right].
\end{equation}
Thus, the Green function is expressed in terms of the single function $S$ defined by (\ref{1-6.1}).

This expression is valid even when there are bound states.
The right-hand side of (\ref{1-6.2}) becomes infinite when $S(x;k)=1$. 
Note that $S(x;k)=1$ is equivalent to $R_l(\infty,x;k)R_r(x,-\infty;k)=1$, which is obviously the condition for resonance. 
(This condition does not depend on $x$; if $S(x;k)=1$, then $S(x';k)=1$, too.)   
This means that $k^2$ is an eigenvalue of the Schr\"odinger operator if $S(x;k)=1$.
Otherwise, (\ref{1-6.2}) is always finite~\footnote{
If $f(+\infty)$ and $f(-\infty)$ are both finite, or if ${\rm Im}\,k>0$, then $\vert S(x;k) \vert$ is finite for any $x$. In other cases it may happen that $\vert S(x;k) \vert=\infty$ for some $x$, but this causes no problems.
}.

\section{Relation with the WKB approximation}
It is interesting to think about the connection between (\ref{1-6.2}) and the WKB method.
In the leading order WKB approximation, a wave function satisfying (\ref{1-1.1}) has the form
\begin{equation}
\label{2-7.1}
\psi(x) \simeq \frac{1}{\sqrt{p(x)}}
\exp\left[i \int^x p(z) \,dz \right],
\end{equation}
where $p(x)$ is the local wavelength for the Schr\"odinger equation defined by
\begin{equation}
\label{2-7.2}
p(x) \equiv \sqrt{k^2-V_{\rm S}(x)}.
\end{equation}
The function $G(x,x')$ given by (\ref{1-6.2}), with fixed $x'$, is an exact solution of (\ref{1-1.1}) for $x>x'$. Comparing (\ref{2-7.1}) with (\ref{1-6.2}), we can see that the quantity $k(1-S)$ in the exact expression corresponds to the local wavelength $p$ in the WKB approximation. 

Let us first see how (\ref{2-7.1}) can be recovered from (\ref{1-6.2}).
If $f(x)$ is constant, say $f(x)=c$, then $R_r$ and $R_l$ take the form~\cite{theory}
\begin{equation}
\label{2-7.3}
R_r(x,-\infty;k)=-R_l(\infty,x;k)=\frac{ik-i\sqrt{k^2-c^2}}{c}.
\end{equation}
Suppose that, for a non-constant $f(x)$, the reflection coefficients can be approximated by the same form as (\ref{2-7.3}):
\begin{equation}
\label{2-7.4}
\fl
R_r(x,-\infty;k) \simeq \frac{ik-i\sqrt{k^2-f^2(x)}}{f(x)},
\qquad 
R_l(\infty,x;k) \simeq \frac{-ik+i\sqrt{k^2-f^2(x)}}{f(x)}.
\end{equation}
Substituting (\ref{2-7.4}) into (\ref{1-6.1}) gives the approximation for $S$,
\begin{equation}
\label{2-7.5}
S(x;k)\simeq 1-\frac{1}{k} \sqrt{k^2-f^2(x)}.
\end{equation}
If we assume that $f(x)$ varies slowly so that $k^2-f^2 \simeq k^2 -f^2 -f'$, then 
\begin{equation}
\label{2-7.6}
k[1-S(x;k)]\simeq \sqrt{k^2-V_{\rm S}(x)}=p(x).
\end{equation}
This reproduces the WKB approximation (\ref{2-7.1}). 

Let us make a more detailed comparison by considering the higher-order corrections. 
Since the WKB expansion is essentially a high-energy expansion, it can be compared with the expansion in terms of $1/k$. 
The asymptotic expansion of the reflection coefficients in powers of $1/k$ was studied in \cite{analysis}. Using the formulas derived there, we can express the corrections to (\ref{2-7.4}) as a series in powers of $1/k$. We have
\refstepcounter{equation}
\label{2-7.7}
\addtocounter{equation}{-1}
\numparts
\begin{equation}
\fl
R_r(x,-\infty)=
i\frac{k-\sqrt{k^2-f^2}}{f}
-\frac{f'}{(2ik)^2}-\frac{f''}{(2ik)^3} +
\frac{5f^2f'-f'''}{(2ik)^4}+ \cdots,
\end{equation}
\begin{equation}
\fl
R_l(\infty,x)=
-i\frac{k-\sqrt{k^2-f^2}}{f}
-\frac{f'}{(2ik)^2}+\frac{f''}{(2ik)^3} +
\frac{5f^2f'-f'''}{(2ik)^4}+ \cdots.
\end{equation}
\endnumparts
The condition for the validity of (\ref{2-7.7}) as an asymptotic expansion is discussed in~\cite{analysis}.
Substituting (\ref{2-7.7}) into (\ref{1-6.1}), we obtain
\begin{equation}
\label{2-7.8}
\fl
S(x)=1-\frac{\sqrt{k^2-f^2}}{k}+\frac{f'}{2k^2}
+\frac{1}{8k^4}\left[2f^2f'-(f')^2-2ff''-f'''\right]+\cdots.
\end{equation}
(A formula is available for the coefficient of $1/k^n$ in the expansion (\ref{2-7.8}) for an arbitrary positive integer $n$.)
By using
\begin{equation}
\label{2-7.9}
\sqrt{k^2-V_{\rm S}}
=\sqrt{k^2-f^2}-\frac{f'}{2k}-\frac{(f')^2+2f^2f'}{8k^3}+ \cdots,
\end{equation}
we can rewrite (\ref{2-7.8}) in terms of $V_{\rm S}$ as
\begin{equation}
\label{2-7.10}
\fl
S(x)=1-\frac{\sqrt{k^2-V_{\rm S}}}{k}-\frac{V_{\rm S}''}{8k^4}
-\frac{1}{32k^6}\left[5 (V_{\rm S}')^2+6 V_{\rm S}V_{\rm S}''-V_{\rm S}^{(4)}\right]+\cdots.
\end{equation}
In the WKB method, on the other hand, the wave function incorporating the higher-order corrections is written as \cite{bender}
\begin{equation}
\label{2-7.11}
\psi(x)=\frac{1}{\sqrt{W(x)}} \exp\left[i\int^xW(z)\,dz\right],
\end{equation}
\begin{equation}
\label{2-7.12}
W = W_0 + W_1 + W_2 + \cdots,
\end{equation}
where
\begin{equation}
\label{2-7.13}
\fl
W_0(x)=p(x), \qquad W_1(x)=-\frac{p''}{4p^2}+\frac{3(p')^2}{8p^3}, \qquad
W_2(x)=\frac{1}{16}\frac{p^{(4)}}{p^4}+ \cdots, \quad {\rm etc.}
\end{equation}
The WKB expansion (\ref{2-7.12}) is an expansion in powers of the constant $\hbar^2$ (which we have set to be unity) which multiplies the fist term on the left-hand side of (\ref{1-1.1}).
It is easy to see that $W_1=O(1/k^3)$, $W_2=O(1/k^5)$, etc as $k \to \infty$. 
(The terms of $W_2$ omitted in (\ref{2-7.13}) are $O(1/k^7)$.) 
So we can rearrange (\ref{2-7.12}) into an expansion in powers of $1/k$. 
We have
\begin{equation}
\label{2-7.14}
\fl
W_1=\frac{V_{\rm S}''}{8k^3}+\frac{3V_{\rm S}V_{\rm S}''}{16k^5}+\frac{5(V_{\rm S}')^2}{32k^5}+O(1/k^7), \qquad
W_2=-\frac{V_{\rm S}^{(4)}}{32k^5}+O(1/k^7), \quad {\rm etc.}
\end{equation}
It is obvious that (\ref{2-7.10}) is equivalent to (\ref{2-7.12}) with (\ref{2-7.14}), where $W=1-kS$. 

Let us next see how the Bremmer series can be described in our formalism.
For this purpose, it is convenient to make use of the rotation introduced in section~5 with an imaginary angle $\theta$. 
We define
\begin{equation}
\label{2-7.15}
\theta(x)\equiv \arctan\frac{if(x)}{k}.
\end{equation}
If $f$ is a constant, the rotation (\ref{1-5.3}) with this angle $\theta$ is a transformation to the frame of coordinates in which no scattering takes place~\cite{algebraic}. 
Using this $\theta(x)$, the quantity on the right-hand sides of (\ref{2-7.4}) can be written as
\begin{equation}
\label{2-7.16}
i \frac{k-\sqrt{k^2-f^2(x)}}{f(x)}=\tan \frac{\theta(x)}{2}.
\end{equation}
The expressions (\ref{2-7.4}) are exact if $\theta(x)$ is an $x$-independent constant.
The corrections to (\ref{2-7.4}) can be expressed as a series in powers of $\theta'(x)=(d/dx)\theta(x)$:
\refstepcounter{equation}
\label{2-7.17}
\addtocounter{equation}{-1}
\numparts
\begin{eqnarray}
\fl
R_r(x,-\infty)&=
\tan \frac{\theta(x)}{2}
-\sec^2 \frac{\theta(x)}{2}\
\Biggl[
\frac{1}{2} \int_{-\infty}^x dz
\,\theta'(z)  e^{2iA_1}
\nonumber \\
\fl
& \quad 
+\frac{1}{8}
\int_{-\infty}^x dz_1\,\int_{-\infty}^{z_1} dz_2\,\int_{-\infty}^{z_1} dz_3\,
\theta'(z_1) \theta'(z_2) \theta'(z_3) e^{iA_2}+ \cdots
\Biggr],
\end{eqnarray}
\begin{eqnarray}
\fl
R_l(\infty,x)&=
-\tan \frac{\theta(x)}{2}
-\sec^2 \frac{\theta(x)}{2}
\Biggr[
\frac{1}{2} \int_x^\infty dz
\,\theta'(z) e^{-2iA_1}
\nonumber \\
\fl
& \quad 
+\frac{1}{8}
\int_x^\infty dz_1\,\int_{z_1}^\infty dz_2\,\int_{z_1}^\infty dz_3\,
\theta'(z_1) \theta'(z_2) \theta'(z_3) e^{-iA_2}+ \cdots
\Biggr],
\end{eqnarray}
\endnumparts
where we have defined
\begin{eqnarray}
\label{2-7.18}
q(x) &\equiv \sqrt{k^2-f^2(x)},
\end{eqnarray}
\numparts
\begin{eqnarray}
A_1 &\equiv \int_z^x q(w)\,dw, 
\\
A_2 &\equiv \int_{z_2}^x q(w)\,dw +\int_{z_3}^x q(w)\,dw 
+\int_{z_2}^{z_1} q(w)\,dw+\int_{z_3}^{z_1} q(w)\,dw.
\end{eqnarray}
\endnumparts
(Since it is not the purpose of the present paper to discuss the approximation methods for the reflection coefficients, we omit the explanation here. Let us only mention that (\ref{2-7.17}) can be derived from equations (10.2) and (10.3) of \cite{algebraic}.)
From (\ref{2-7.17}) and (\ref{1-6.1}) we obtain
\begin{eqnarray}
\label{2-7.20}
\fl
S(x)&=
1-\frac{q(x)}{k}
-\frac{1}{2\left[\cos \frac{1}{2}\theta(x)+\sin \frac{1}{2}\theta(x)\right]^2}
\int_{-\infty}^x dz
\,\theta'(z) \exp\left[2i\int_z^x q(w)\,dw \right]
\nonumber \\
\fl 
& \quad
-\frac{1}{2\left[\cos \frac{1}{2}\theta(x)-\sin \frac{1}{2}\theta(x)\right]^2}
\int_x^\infty dz
\,\theta'(z) \exp\left[2i\int_x^z q(w)\,dw \right]+\cdots.
\end{eqnarray}
(It is possible to construct a formula for the term of an arbitrary order in the expansion (\ref{2-7.20}).) 
Just like (\ref{2-7.10}), we can rewrite (\ref{2-7.20}) in terms of $p$ as
\begin{eqnarray}
\label{2-7.21}
S(x)&=
1-\frac{p(x)}{k}
+ \frac{p(x)}{2k}\int_{-\infty}^x dz
\, \frac{p'(z)}{p(z)}\exp\left[2i\int_z^x p(w)\,dw \right]
\nonumber \\
& \quad
- \frac{p(x)}{2k}\int_x^\infty dz
\, \frac{p'(z)}{p(z)}\exp\left[2i\int_x^z p(w)\,dw \right]+\cdots.
\end{eqnarray}
Substituting (\ref{2-7.21}) into ({\ref{1-6.2}), expanding the right-hand side, and carrying out integration by parts, we obtain
\begin{eqnarray}
\label{2-7.22}
\fl
G(x,x')&=
\frac{1}{\sqrt{p(x)p(x')}}\exp\left[i\int_{x'}^xp(z)\,dz\right]
\Biggl\{
1+\frac{1}{2}\int_{-\infty}^{x'}dz
\, \frac{p'(z)}{p(z)}\exp\left[2i\int_z^{x'} p(w)\,dw \right]
\nonumber \\
\fl
& \quad  \qquad \qquad 
- \frac{1}{2}\int_x^\infty dz
\, \frac{p'(z)}{p(z)}\exp\left[2i\int_x^z p(w)\,dw \right]+\cdots
\Biggr\},
\end{eqnarray}
which corresponds to the ordinary Bremmer series.
Whereas the Bremmer series is an expansion of the wave function $\psi(x)$, the expression (\ref{2-7.20}) or (\ref{2-7.21}) gives a similar expansion for the quantity corresponding to $W(x)$ of equation~(\ref{2-7.11}).

As we have noted, the WKB approximation (\ref{2-7.1}) is obtained by replacing the Fokker-Planck potential $V(x)$ by a linear function at each point $x$. 
Another possible approximation is to replace $V(x)$ by a quadratic function at each $x$. 
The reflection coefficients for quadratic potentials can be exactly obtained \cite{analysis}. 
By substituting these exact expressions into (\ref{1-6.2}) with (\ref{1-6.1}), we obtain an approximation for the Green function.  
In some cases, this approximation can be better than the WKB approximation.

The methods related to the WKB approximation we have seen above is just an example of using (\ref{1-6.2}) for approximate evaluation. In making an approximation, in general, it is easier to deal with the reflection coefficients than the Green function itself. 
For each approximation method for the reflection coefficients, the expression (\ref{1-6.2}) gives the corresponding approximation for the Green function.

\section{Conclusion}
In this paper, we have derived some exact expressions for the Green function.
A general symmetric expression is given by (\ref{1-5.32}).
Reflecting the symmetry of the Fokker-Planck equation, this expression includes an arbitrary parameter $\eta$. 
The simplest expression (\ref{1-6.2}) is obtained by setting $\eta=1$. 
Analytic properties of the reflection coefficients can be studied relatively easily. By using the expressions derived here, we can investigate the properties of the Green function on the basis of the analysis of the reflection coefficients. In particular, (\ref{1-6.2}) is useful for studying the high-energy behavior of the Green function. It also serves as a starting point for various approximation methods.

The reflection coefficients $R_r(x,-\infty)$ and $R_l(\infty,x)$ that appear in our expressions have been defined by using the Fokker-Planck equation.
Of course, this is not the only possible way of defining reflection coefficients for semi-infinite intervals. 
It is also possible to express the Green function in terms of reflection coefficients defined in a different way, without using the Fokker-Planck equation.
However, the resulting expressions become more complicated if we use a different (inequivalent) definition of the reflection coefficients.

%%%%%%%%%%%%%%%%%%%%%%%%%%%%%%%%%%%%%%%%%%%%%%%%%%%%%%%%%%%

%%%%%%%%%%%%%%%%%%%%%%%%%%%%%%%%%%%%%%%%%%%%%%%%%%%%%%%%%%%

\end{document}